# A Secure TFTP Protocol with Security Proofs

Mohd Anuar Mat Isa[1], Habibah Hashim[2], Syed Farid Syed Adnan[3], Jamalul-lail Ab Manan[4], Ramlan Mahmod[5]

*Abstract*— Advances in smart devices has witnessed major developments in many mobile applications such as Android applications. These smart devices normally interconnect to the internet using wireless technology and applications using the TFTP protocol among these wireless devices are becoming commonplace. In this work, we present an enhanced lightweight security protocol for smart device and server communications using Trivial File Transfer Protocol (TFTP). We suggest the use of lightweight symmetric encryption for data encryption and asymmetric encryption for key exchange protocols in TFTP. The target implementation of secure TFTP is for embedded devices such as Wi-Fi Access Points (AP) and remote Base Stations (BS). In this paper we present the security proofs based on an attack model (IND-CCA2) for securing TFTP protocol. We also present the security reduction of SSW-ARQ protocol from Cramer-Shoup encryption scheme and *fixed-time* side channel security. We have also introduced a novel adversary model in IND-CCA2-(SC-TA) and it is considered a practical model because the model incorporates the *timing attack*.

*Index Terms*— Cryptography, TFTP, IND-CCA2, Timing Attack, Cramer Shoup, Stop and Wait ARQ, Smart Environment, Trivial File Transfer Protocol, Wi-Fi AP, Security, Trust, Privacy, STP, Trusted Computing, UBOOT, AES, IOT, Access Point, AP, Base Station, BS, WIFI, UDP, Lightweight, Asymmetric, Symmetric, Reductionist

## I. INTRODUCTION

THIS paper is a continuation from our previous work. Related works with regard to improvements in the TFTP protocol had been quiet for almost 10 years. The most recent publication was in RFC 3617 (2003) [1]. The RFC 3617 mentioned that there is *"no mechanism for access control within the protocol, and there is no protection from a man in the middle attack"*. Our publication in 2013 [2] proposed an implementation of a lightweight and secure TFTP protocol for embedded systems. We proposed a new packet header for RRQ, WRQ and OACK. These headers provide security information for TFTP's data payload encryption. However, we did not discuss about the implementation, confidentiality, integrity, authenticity and the attack model that could compromise the new proposed

Manuscript received March 14, 2014; revised April 10, 2014. The authors would like to thank to Ministry of Higher Education (MOHE) for providing the grant 600-RMI/NRGS 5/3 (5/2013), and Universiti Teknologi MARA (UITM) for providing the research grant 600-RMI/PSI 5/3.

Faculty of Electrical Engineering, 40450 UiTM Shah Alam, Selangor, Malaysia.     [1]anuarls@hotmail.com,     [2]habib350@salam.uitm.edu.my (corresponding author), [3]syed_farid@salam.uitm.edu.my
MIMOS Berhad, Technology Park Malaysia, 57000 Kuala Lumpur, Malaysia. [4]jamalul.lail@mimos.my
Faculty of Computer Science & Information Technology, 43400 Universiti Putra Malaysia, Serdang, Selangor, Malaysia. [5]ramlan@upm.edu.my

TFTP protocol. Also missing was the role of Message Authentication Code (MAC) in the overall scheme. The MAC must be used to ensure encrypted TFTP data payload is unchanged by attackers or transmission bit errors.

After last year's publication, it was thought that there is no interest from others to use or explore this protocol. However, when we checked our personal account in the *Academia.edu* in the *Analytics* section, we found that almost everyday the paper [2] was hit by the search engine for almost six months. Recently, we received an email that requested advice for a lightweight TFTP protocol in cloud computing. We take this as a sign that we need to further explore to enhance the TFTP lightweight security scheme. This motivates us to continue the research and thus publish this paper.

This paper was written in a general information security terminology with a simple mathematical notation (semi-formal). It is intended for information security practitioners and not for mathematicians or cryptographers as the main audience. We hope that this paper will give a worthy understanding of cryptographic scheme and its security proofs. We also understand that it was tough for a non-mathematical background to grasp the *reductionist style*. Therefore, In this paper we taken a simplistic approach and we have skipped the math intensive parts in the Sections *V: Security Property* and *VI: Security Analysis* which can be obtained from references [3–5]. We hope that, with this approach, the reader can easily understand the security proofs presented for the TFTP lightweight security scheme in designing or implementing a networking protocol or application.

## II. RESEARCH GOAL

### A. Objective

The purpose of this research work is to facilitate security in the TFTP protocol. We introduced Cramer-Shoup[3] encryption scheme and *fixed-time* side channel security as underlying security protocol for a new secure TFTP.

### B. Motivation

Referring to our previous work [2], we have mentioned the need of a secure TFTP protocol particularly in various network administrative tasks such as monitoring and upgrading of remote embedded device's firmware, where a lightweight protocol such as TFTP is usually employed. The security risks in such situations were also discussed with emphasis on concerns due to physical attacks, wherein attackers access and modify Wi-Fi AP hardware and software [2], [6], [7]. In a preceding work, we proposed an enhanced data communication package for DENX-UBOOT [8] firmware to include a secure TFTP protocol. However, our proposal did not suggest a specific cryptographic protocol for the successful implementation of the secure TFTP protocol. In the effort to further augment the work, a proven secure and practical asymmetric cryptographic scheme, i.e. the Cramer-Shoup (CS) protocol is proposed to be deployed as the underlying cryptographic protocol [3] in the overall scheme. . In the latter part, the CS will provide a secure asymmetric key exchange, wherein CS will be used to encrypt symmetric key (e.g., AES 512) for a secure TFTP data communication.

### III. NOTATION AND DESCRIPTION

#### A. Operator

*a) Modular Arithmetic (Congruence)[1]*

$$x \equiv y \ (mod \ p) \ where \ x, y \ \in integers \ and$$

$$p \in positive \ integer$$

Therefore, $x$ is congruent to $y$ modulo $p$ (or $x$ is residue of $y$ modulo $p$). $E.g. \ 3^2 \ (mod \ 7)$

$$9 \equiv 2 \equiv -5 \ (mod \ 7)$$

*b) Primitive Root*

$$Let \ g^x \in \mathbb{Z}_p \ where \ p \ is \ a \ prime$$

$$if \ p = 7, x \in (1 \dots p - 1) \ then \ g^x (mod \ p) \equiv ?$$

TABLE I
Primitive Root for Generator $g$

| x | 1 | 2 | 3 | 4 | 5 | 6 | Sorted Result |
|---|---|---|---|---|---|---|---|
| g = 3 | 3 | 2 | 6 | 4 | 5 | 1 | 1, 2, 3, 4, 5, 6 |
| g = 4 | 4 | 2 | 1 | 4 | 2 | 1 | 1, 1, 2, 2, 4, 4 |

g = 3: complete elements of $\{1,2,3,4,5,6\} \in (1..p-1) \ or \ \mathbb{Z}_p$

$\therefore g = 3$ is a primitve root of 7 (generator[2] 3 for $\mathbb{Z}_7$)

$$\mathbb{Z}_p = \mathbb{Z}_7 = \{3^1, 3^2, 3^3, 3^4, 3^5, 3^6\}$$

$\therefore$ a prime order p of cyclic group G is $|\mathbb{Z}_p| = |\mathbb{Z}_7| = p - 1 = 6$

However, $g = 4$ is not primitive root of $\mathbb{Z}_7$

#### B. Reduction

The reduction approach can show that hardness (difficulty or intractable) of one problem $P1$ implies hardness of another problem $P2$ given that $P2$ has been reduced to $P1$. By security reduction, we consider that if someone has an algorithm $A1$ that can solve a computationally hard problem $P2$, then if the same algorithm $A1$ with a little modification can also solve $P1$, then we can conclude that problem $P1$ has been reduced to problem $P2$ with notation $P1 \leq P2$ [9]. The reduction technique was used in the NP-completeness theory [10] to prove the NP-completeness of a problem such that if $P1$ is NP-complete problem and $P2$ is another NP problem; then it can prove that $P2$ is also an NP-complete problem, if $P1 \leq P2$.

### IV. RELATED WORK

#### A. Trivial File Transfer Protocol (TFTP)

TFTP is a simple protocol that has been widely used for transmitting files albeit with limited functionalities [11]. It provides upload and download operations using UDP protocol. The actual transmission protocol that is used to control file transfer is *"Simplex Stop and Wait with Automatic Repeat reQuest"* (SSW-ARQ). TFTP was designed as an application for the Internet Protocol (IP) [12] because at that moment, computers or embedded systems do not have sufficient memory or lack disk space to provide full FTP support. Nowadays, TFTP is quite popular and it is used by network administrators to upgrade router firmware and to distribute software within a corporate network (e.g., DENXU-Boot [8] firmware). Thus, it is beneficial for booting embedded devices (e.g., sensor nodes) that may not have sufficient volatile memory to store OS kernel and applications.

Recently, there have been some research works which have addressed the potential usage of TFTP protocol for Radio Frequency (RF) [12], remote attestation for Trusted Computing [13] (e.g., Trusted Platform Modules (TPM)), lightweight protocol for remote accessing the cloud infrastructure [14], Wide Area Network (WAN) surveillance system [7], [15] and etc. However, their suggestions to use TFTP as medium in their research frameworks are not practical and not secure mainly because TFTP exposes all data packet in plaintext. The authors should not assume that TFTP can provide secure communication (confidentiality, integrity and authenticity) for data transfer.

#### B. Simplex Stop and Wait Automatic Repeat Request (SSW-ARQ)

SSW-ARQ is a simple network protocol used by network applications (e.g., TFTP) to enable stop and wait flow control in frame transmission when using unreliable UDP/IP stacks [11], [16]. It allows retransmission of frames in the event of frame loss or corrupted frame [17][11]. Fig. 1 shows an example of frame transmission using SSW-ARQ. To enable security in this protocol, we may integrate it with Cramer-Shoup[3] encryption scheme in the frame data payload.

From Fig. 1, **A** wants to transmit data or file to **B** in a secure manner. Therefore, both parties need to establish a secure key exchange for symmetric encryption (e.g., share AES512's secret keys). Before that, the AES512's secret keys must be shared in a secure communication protocol and this can be accomplished using Cramer-Shoup[3] encryption scheme. In this communication setup, both parties are pre-installed with Cramer-Shoup's asymmetric keys by the network administrator before this communication happen. It is assumed that both parties who are communicating with each other are in full knowledge of the recipient's public key.

---

[1] Modular arithmetic operation is based on set elements in a finite abeliangroup $\mathbb{G}$. One can refer a book *"Introduction to modern cryptography"* [5] for further crypto discussion. The book provides a good explanation for non-crypto reader.
[2] It is also called a *"cyclic group"* wherein all elements in the group are generated using single element such as generator $g$.

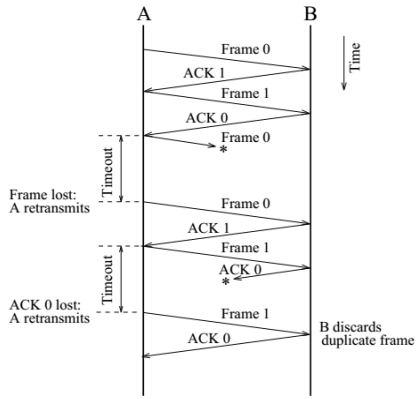

Fig. 1. SSW-ARQ protocol[18]

The communication begins with **B** who generates the AES512's secret keys. Then, the AES512's secret keys is wrapped (encrypted) using **B**'s public key. Due to limitation of SSW-ARQ's frame size, a ciphertext generated using **B**'s public key must be divided into chunks that fit into the frame. After that, **A** will transmit multiple frame segments containing the chunks of ciphertext. However, the SSW-ARQ communication protocol allows only one frame to be sent at one time. The next frame will be transmitted after receiving a correct acknowledgement (ACK) from **B**. At this stage, all transmitted frame must verify that it is free from data corruption (e.g., bit-error) using the checksum function. After all frames has been successfully transmitted, **B** will assemble all frame segments into the complete ciphertext string. After that, **B** will call Cramer-Shoup[3] decryption function to decrypt the ciphertext and then retrieve the AES512's secret keys. Finally, **A** will encrypt the file using the AES512's secret keys and send the encrypted file using standard TFTP protocol. B will decrypt the file using the AES512's secret keys. However, in this paper, we will not discuss the usage of symmetric encryption scheme and its security.

slower than the El-Gamal (approximately twice) in performing cryptographic computation [20]. To compare against RSA, Cramer-Shoup is slower in the encryption process but it is slightly equal in the decryption process [20]. We illustrate the Cramer-Shoup protocol in Fig. 2.

## V. SECURITY PROPERTY

### A. IND-CCA2

Indistinguishability-Adaptive Chosen-Ciphertext Attack [21] is an attack that allows an adversary to access a decryption function through the decryption oracle. The adversary can ask the oracle to decrypt any ciphertext except the one that being use for indistinguishability test. The IND-CCA2 allows the Adversary to get a decryption of ciphertext from the oracle in Phase 1(before) and Phase 2 (after) the challenge messages ($m_0, m_1$ where $|m_0| = |m_1|$) are issued to Challenger.

For the indistinguishability test, the adversary will send two plaintext messages ($m_0, m_1$) to the Challenger. In place of a fair indistinguishability experiment, both plaintext messages must never be used for decryption using the oracle. This means that the adversary could never know the ciphertext of both messages after the encryption function has been applied. Referring to Fig. 3, the Challenger will choose randomly either $m_0$ or $m_1$ to be encrypted. Ciphertext $c$ of the encrypted message $m_0$ or $m_1$ is sent to the Adversary. The Adversary need to distinguish the whether the ciphertext $c$ is either $m_0$ or $m_1$ with probability of $\frac{1}{2}$. If the probability to guess a correct the ciphertext c is greater than $\frac{1}{2}$, we can conclude that the Adversary has an *"advantage"* and the given protocol is considered not secure in terms of indistinguishability.

*Let $p(n)$ denote the set of prime in size of $n$,*

*for all sufficiently large n in $IND - CCA2$.*

$$|Pr[success] - Pr[failure]| < \frac{1}{p(n)}$$

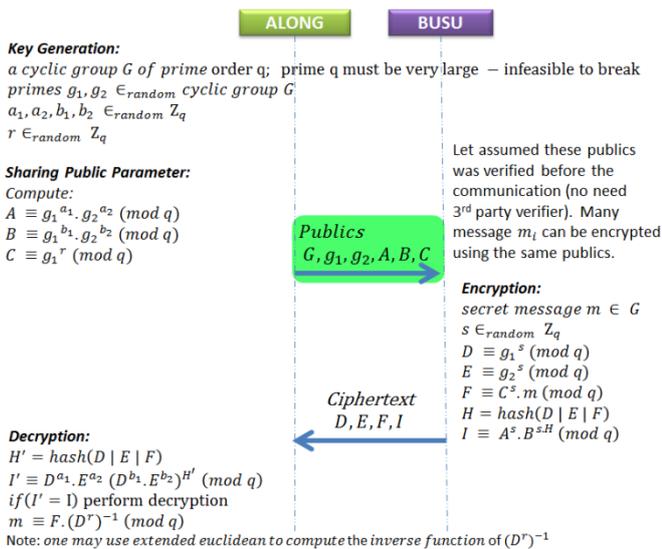

Fig. 2. A simplified Cramer-Shoup Encryption Scheme

### C. Cramer-Shoup Encryption Scheme

Cramer-Shoup[3] protocol is proven secure against IND-CCA2. The protocol provides an improvement of El-Gamal[19] wherein the El-Gamal is vulnerable to chosen-ciphertext attack (CCA). However, the Cramer-Shoup is

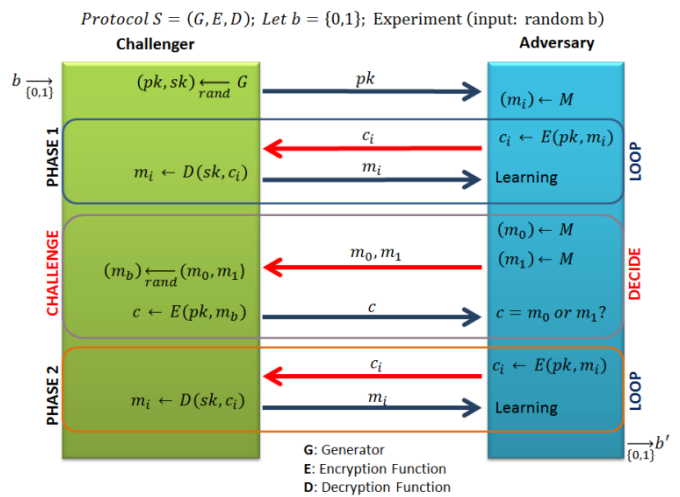

Fig. 3. IND-CCA2's Experiment

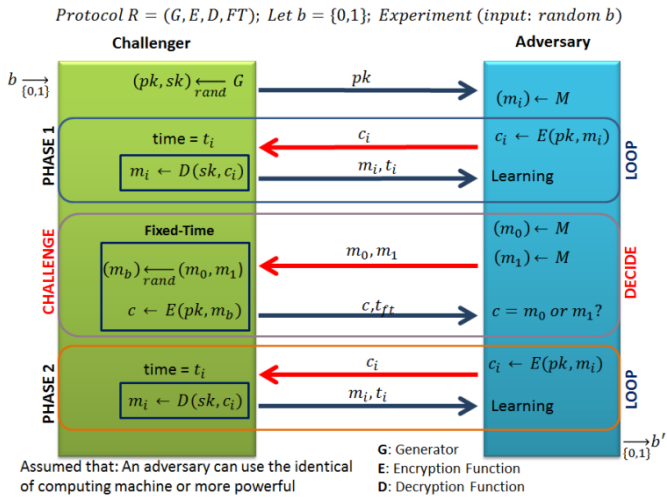

Fig. 4. IND-CCA2-(SC-TA)'s Experiment

### B. IND-CCA2-(SC-TA)

Indistinguishability-Adaptive Chosen-Ciphertext Attack-(Side Channel – Timing Attack) is an attack that allows an adversary to access identical computing resources in terms of computing power (e.g., CPU). The adversary is given knowledge of time to perform cryptographic computations (e.g., primitive computation and protocol execution). These were included given that the adversary has knowledge of the delay of network transmission for all transactions in Phase 1, Phase 2 and Challenge phase (refer to Fig. 4). The adversary also has the knowledge of IND-CCA2 given that the Adversary's *"advantage"* over random guessing in indistinguishability test with Timing-Attack is:

Let $p(n)$ denote the set of prime in size of $n$,

for all sufficiently large $n$ in $IND - CCA2 - (SC - TA)$.

$$|Pr[success] - Pr[failure]| < \frac{1}{p(n)}$$

## VI. SECURITY ANALYSIS

### A. Cramer-Shoup with IND-CCA2

**Adversary Model:** Adaptive Chosen-Ciphertext Attack (CCA2).

**Security Claim:**

*1.1)* Decision Diffie-Hellman Problem (DDHP) problem is hard [4] in a cyclic group $G$;

*1.2)* Hash function is a universal one-way hash function with strong collision-resistant [3], [22]; Then, Cramer-Shoup encryption scheme is secure against CCA2 using indistinguishability test.

**Security Reduction:** An adversary claims that he can break Cramer-Shoup protocol using an efficient algorithm $A$ in a program $A$. To test the adversary claim, we conduct an experiment by taking the program $A$ and put a simple "*wrapper*" into it, and we call it program $A'$. The program $A'$ will use the program $A$ as a sub-routine in the experiment. Then, the program $A'$ will run the IND-CCA2 experiment with random input $b$ and with expected output $b'$ in indistinguishability test. The adversary is considered a winner in the experiment, if the probabilities to guess for all correct messages are non-negligible with an *advantage* of $\left(\frac{1}{2}\right) + \varepsilon(n)$, where $\varepsilon(n)$ is the Adversary's success probability. Due to the non-negligible *advantage*, the program $A'$ can break the Cramer-Shoup protocol. However, if there are no other efficient programs (including program $A'$) that can win in the experiment with non-negligible *advantage*, the Cramer-Shoup protocol won the experiment with negligible *advantage* of program $A'$. Since the **Security Claims (1.1 and 1.2)** in the previous paragraph used strong primitive assumptions (DDHP is hard and collision-resistance of hash function), the program $A'$'s *advantage* over probabilistic *polynomial-time*[3] is negligible. Therefore, the program $A'$ lost in the experiment by indistinguishability test with a negligible *advantage* and the adversary claim was invalid (false) in that it *"can break Cramer-Shoup protocol using all efficient algorithm A in a program A"*.

### B. SSW-ARQ with IND-CCA2-(CS-TA)

**Adversary Model:** Adaptive Chosen-Ciphertext Attack-(Side Channel – Timing Attack).

**Security Claim:**

*2.1)* SSW-ARQ inherits all security strength from the Cramer-Shoup encryption scheme and the Cramer-Shoup encryption scheme was proven secure in the IND-CCA2.

*2.2)* SSW-ARQ is secure against *Timing Attack* using *fixed-time* of runtime for all fixed input length in the function in a *polynomial time*; in non-formal description: Any same function that receives any valid input with the same length (e.g., $f(101)$ and $f(001)$, where $|f(101)| = |f(001)|$) will have identical runtime or execution for all conditions; Then, SSW-ARQ protocol is secure against CCA2-(CS-TA) using indistinguishability test.

**Security Reduction:** For the **Security Claim 2.1)**, it was easy to observe the security proof because Cramer-Shoup encryption scheme was embedded into SSW-ARQ protocol. All strings (e.g., ciphertext, public key) that are generated by Cramer-Shoup encryption scheme are divided into chunks that are fitted into the SSW-ARQ's frame. Any modification (even a single bit error) in the SSW-ARQ's frame will result in a failure in Message Authentication Codes (MAC) in the Cramer-Shoup encryption scheme. This good security property was derived from the collision-resistant hash function. Therefore, *"Given that Security Claim **2.1** is true, the SSW-ARQ is secure against IND-CCA2"*.

For the **Security Claim 2.2)**, we can use a similar experiment that is used for Cramer-Shoup encryption scheme except that an adversary are given knowledge of runtime performance of cryptographic computation and network transmission delay.

Referring to **Security Claim 2.2**, it is impossible to attain the same fixed time for the encryption and decryption process of different input strings of ciphertext (with same length ciphertext and different key) using specific encryption functions or decryption functions. Running time to compute an exponential such as $g^x$ and $g^{x+1}$ is different because of the different computer machine capabilities in performing addition to representing multiplication as well as the different limitations of hardware data bus. It might be similar for small inputs of 32-bits or 64-bits length, but it is not so for crypto numbers with extensive lengths such as 2048-bits length of public key. From a practical point of view, we can use a subset of the assumption from the **Security Claim 2.2**, *"a fixed-time is based on worst-case scenario to do encryption or decryption process for all string of plaintext or ciphertext that has the same length and within the same cyclic group G of prime order q"* as **Security Claim 2.2.1**. The **Security Claim 2.2.1** show that if we run the IND-CCA2-(SC-TA)'s experiment as shown in the Fig. 4, the program $A'$ was lost in the experiment by indistinguishability test with a negligible *advantage*. This happened because the program $A'$ cannot distinguish whether the ciphertext $c$ was either $m_0$ or $m_1$ with a given worst case fixed-time. For example that based on Fig. 4, if a given message size of $m_0 = \{0\}^{2048}$ and $m_1 = \{1\}^{2048}$, and the encryption function always gives worst case time, $t_{ft} = (time(Enc(pk, m_b)) + delay)$. The probability to guess a correct message by program $A'$ is $(\frac{1}{2})$ for either $m_0$ and $m_1$:

$$t_{ft} = (time(Enc(pk, m_o) + delay_j)$$
$$t_{ft} = (time(Enc(pk, m_1) + delay_k)$$

---

[3] "polynomial-time" is a term used for measuring an algorithm's running time as a function, wherein it is measured by length of its input into the function [5]. E.g. function $f(x)$ take $x = 1024$ as input string during execution, then the running time is $x$.

The program $A'$ needs to distinguish the ciphertext $c$ through the timing knowledge of time $t_{ft}$. However, the program $A'$'s knowledge of time $t_i$ from the oracle in Phase 1(before) and Phase 2 (after) is not helpful to give non-negligible advantage in the indistinguishability test. Since the **Security Claims (2.1 and 2.2.1)** in the previous paragraph used the Cramer-Shoup encryption scheme, and the *fixed-time* (worst-case scenario) security assumptions, thus the program $A'$'s *advantage* over probabilistic *polynomial-time* is negligible. Therefore, the program $A'$ lost in the experiment by indistinguishability test with a negligible *advantage* and the adversary claim was invalid (false) in that it *"can break the new fixed-time SSW-ARQ protocol (with the IND-CCA2-(SC-TA) attack model) using all efficient algorithm A in a program A"*.

VII. DISCUSSION

We propose to implement security in the TFTP protocol. Sections V and VI has discussed the security properties and security proofs with a strong assumptions of cryptographic primitive. Both sections only showed the security of SSW-ARQ protocol against IND-CCA2-(SC-TA) but not the TFTP protocol wherein the SSW-ARQ protocol is a subset of the TFTP protocol. In our case, TFTP is just an application that manages file transfer and key management. The TFTP will invoke the file transfer using SSW-ARQ protocol and passes a security related key that is needed by SSW-ARQ protocol to perform cryptographic computation (e.g., Cramer-Shoup protocol). Therefore, to prove that the TFTP application is secure, the TFTP must be programmed to follow the standard [23], [24] and practice [25] for a secure application. However, this is beyond the scope of this research paper.

A secure key management protocol in the TFTP application plays an important role to ensure all cryptographic schemes are secure. Bad implementation of key management will expose the cryptographic scheme through many side-channel attacks such as timing attacks, power monitoring attacks and etc. These security vulnerabilities can be exploited in generating, distributing and managing cryptographic keys for embedded devices (e.g., RaspberryPi board) and DENX-UBOOT's TFTP application. Tamper resistant devices can be integrated into embedded hardware for protecting the cryptographic keys such as TPM chip [26]. To minimize our research scope, we have not considered the physical security attacks and the side-channel attacks except for timing attacks in TFTP.

We have introduced a novel adversary model in IND-CCA2-(SC-TA). This adversary model includes knowledge of time to perform cryptographic computation. This makes the Adversary become more powerful than adversary model in IND-CCA2. For example, if the *timing attack* is mounted into the IND-CCA2, the Adversary has a significance non-negligible advantage. The Adversary can build a timing dictionary for every request of decryption of ciphertext $c_i$ with time $t_i$ in Phase 1 and Phase 2. The timing dictionary will give a non-negligible advantage to the Adversary to choose a correct encrypted message by a given ciphertext $c$ in the Challenge process.

However, the timing dictionary for the IND-CCA2-(SC-TA) is unable to choose the correct encrypted message because of *fixed-time* constraint in $(m_b, c, t_{ft})$. We believe that, the IND-CCA2-(SC-TA)'s adversary model will provide a sufficient proof to assert that SSW-ARQ protocol is secure in the indistinguishability test and secure in timing attack. The *fixed-time* using "*worst-case scenario*" is a practical solution to be implemented in the DENX-UBOOT's TFTP application. One may think that using "*worst-case scenario*" slows down the security computation but based on observations in our laboratory, to transmit a file (e.g., Linux Kernel "*wheezy-raspbian*"[27] 2.8MB size) using DENX-UBOOT's TFTP application; the required Estimated Time of Completion (ETC) is around 15-30 seconds. Adding an extra 3-7 seconds to implement the security protocol in the DENX-UBOOT's TFTP application can be considered quite negligible.

VIII. CONTRIBUTION

The overall view of this paper and its contributions were mapped in the Fig. 5. Based on our current and previous effort [2], [6], [7], we have discussed a security framework, method and protocol which would secure TFTP communication. In this paper, we are focused on proving that the enhanced TFTP protocol is secure using a semi-formal notation and reduction technique. The security proofs of TFTP protocol that is given by us can be used in Common Criteria's Evaluation Assurance Level 6 (EAL6) [24]. The EAL6 accept a semi-formal verified design and security test for a target system (e.g., secure TFTP). We have performed a security analysis and demonstrated that the enhanced TFTP is resistant to attacker penetrations related to IND-CCA2 and IND-CCA2-(SC-TA). We have also introduced a novel adversary model in IND-CCA2-(SC-TA) and it is a practical model used to test resistance against timing-attack. For an implementation of secure TFTP, we have provided the proofs and the practical implementation of this new protocol can be initiated. A proper implementation of secure TFTP will ensure remote system updating and patching (e.g., firmware, kernel or application) are secure from attempts to eavesdrop and modify the TFTP's packet.

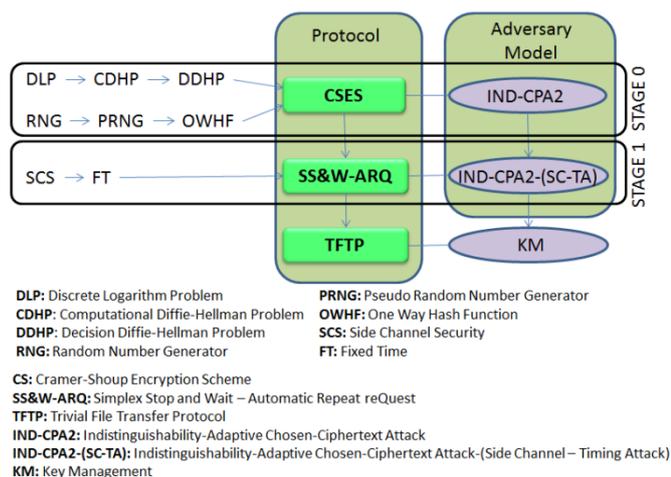

Fig. 5. Summary of security protocol with adversary model

## IX. CONCLUSION

In this paper, we presented the security proof and an attack model for a secure TFTP protocol. We also presented the security reduction of SSW-ARQ protocol from Cramer-Shoup encryption scheme and *fixed-time* side channel security. The secure TFTP protocol would overcome security problems (confidentiality, integrity and authenticity) in controlling, monitoring and upgrading embedded infrastructure in a pervasive computing environment. The target implementation of secure TFTP is for embedded devices such as Wi-Fi Access Points (AP), remote Base Stations (BS) and wireless sensor nodes. In the next stage of our research work, we want to implement a secure TFTP in radio frequency (RF) communication for Structural Health Monitoring (SHM) in electrical pylon tower.